\documentclass[showpacs,preprintnumbers,amsmath,amssymb]{revtex4}
\usepackage{graphicx}
\usepackage{dcolumn}
\usepackage{bm}

\begin{document}

\title {Low Temperature Transport and Specific Heat Studies of Nd$_{1-x}$Pb$_{x}$MnO$_{3}$ Single Crystals}


\author{N.Ghosh$^{a,*}$,U.K. R{\"o}{\ss}ler$^{b}$, K. Nenkov$^{b}$,C. Hucho$^{a}$, H.L.Bhat$^{c}$ and  K.-H.M{\"u}ller$^{b}$\\
{\small $^{a}$ Paul Drude Institut f{\"u}r Festk{\"o}rperelektronik, Hausvogtei Platz 5-7, Berlin-10117, Germany}\\
{\small $^{b}$ IFW Dresden, P.O.B 270116, 01171 Dresden, Germany}\\
{\small $^{c}$ Physics Department, Indian Institute of Science, C.V. Raman Avenue, Bangalore-560012, India}}



\begin{abstract}



Electrical transport and specific heat properties of  Nd$_{1-x}$Pb$_{x}$MnO$_{3}$
single crystals for $0.15 \le x \le 0.5$
have been studied in low temperature regime.
The resistivity in the ferromagnetic insulating (FMI) phase for $x \le 0.3$
has an activated character. The dependence of the activation gap $\Delta$
on doping $x$ has been determined and the critical concentration
for the zero-temperature metal-insulator transition was determined
as $x_{c}$ $\approx$~0.33.
For a metallic sample with $x=$~0.42,
a conventional electron-electron (e-e)
scattering term  $\propto$ $T^{2}$ is found in the low-temperature
electrical resistivity,
although the Kadowaki-Woods ratio is found to be much larger
for this manganite than for a normal metal.
For a metallic sample with $x=$~0.5, a resistivity minimum is
observed for $x$ = 0.5. The effect is attributed to
weak localization and can be described by a
negative $T^{1/2}$ weak-localization contribution to resistivity
for a disordered three-dimensional electron system.
The specific heat data have been fitted to
contributions from free electrons ($\gamma$),
spin excitations ($\beta_{3/2}$), lattice
and a Schottky-like anomaly related to the
rare-earth magnetism of the Nd ions.
The value of $\gamma$ is larger
than for normal metals, which is ascribed to
magnetic ordering effects involving Nd.
Also, the Schottky-like anomaly appears broadened
and weakened suggesting inhomogeneous molecular fields at
the Nd-sites.
\end{abstract}

\maketitle
\section{Introduction}
%
Colossal magnetoresistive manganites (R$_{1-x}$A$_{x}$MnO$_{3}$ )
are known for  interesting physical properties and complicated
electronic phase diagrams \cite{coey, salamon,tokura1, cnr, tokura2}.
In particular, their intrinsic  properties
in the low-temperature phases are far from trivial.
While many of these properties  were known  for more than 50 years
\cite{van}, an appreciation of the magnitude of these effects
is a more recent development \cite{jin}.
Among the different phases in these manganites the occurrence of a
ferromagnetic insulating (FMI) phase poses a difficult problem
because the double exchange model predicts only metallic
ferromagnetism at low temperature \cite{Zener,Anderson,degennes}.
The ferromagnetic insulating behaviour is observed
in manganites when the divalent dopant concentration is quite low.
At higher dopant concentration, crossing the threshold limit,
a ferromagnetic metallic phase (FMM) is usually found.
The low-temperature transport properties of manganites have been
studied in some detail only in recent years \cite{salamon}.
Urushibara {\it et al}.\ described the temperature dependence of
the electrical resistivity in La$_{1-x}$Sr$_{x}$MnO$_{3}$ in
the composition range of metallic conductivity
at low temperature as $\rho$(T)= $\rho$(0)
+ $A\,T^{2}$ \cite{Urushibara}.
Similar data were reported by  Schiffer {\it et al}.\ \cite{Schiffer}
and  Snyder {\it et al}.\ \cite{Snyder}.
As pointed out by Kadowaki and Woods \cite{kadawaki}, a general
relationship between the coefficient $A$ for the $T^{2}$
contribution to the resistivity and the square electronic heat-capacity
coefficient $\gamma$ is obeyed in usual metallic systems.
However, in manganites very large ratios $A/\gamma^{2}$
are found that are
an order of magnitude larger than
the Kadowaki-Woods ratio for metals
with strong electron-electron (e-e) interactions \cite{jaime}.
This casts doubts on the assumption that
the usual electron-electron (e-e) scattering
underlies the $T^{2}$ dependence of
the resistivity observed
in metallic manganites  \cite{salamon}.
In particular, manganites are well known examples for bad metallic
behaviour \cite{Mayr,Riva}.

Some  recent reports find that the manganites have
resistivity minima  and the resistivity at low temperature
is higher than Mott's maximum metallic resistivity of
about 10m$\Omega$ cm \cite{Kumar,Rana,Zhang}.
According to the scaling theory for disordered electronic systems
in 3D metals \cite{Lee}, the  weak localization and
electron-electron scattering in presence of strong disorder should
influence the electrical conduction in manganites.
This type of e-e interaction is described by a negative $T^{1/2}$
power law contribution to the resistivity.
This contribution has been employed
to fit the resistivity data of the intrinsically
disordered systems like manganites in recent
reports \cite{Kumar, Rana, Zhang}.
Among other mechanisms ruling
the temperature dependence of the transport properties,
the contribution of electron-magnon scattering
to the resistivity in manganites poses further problems.
In the ground-state, one-magnon processes should
not take place because of the half-metallic
band-structure of the manganites with
a fully polarized band of the electrons
at the Fermi level \cite{Furu}.
Additionally, the appearance of a Kondo anomaly in this kind of
ferromagnetic compounds with spin-disordered clusters has also been
discussed \cite{Zhang}.
This also suggests that there is a
need of some modification in the dependence
$\rho$(T)= $\rho$(0) + $A\,T^{2}$ to describe
the low temperature metallic
transport property of manganites.

A new theoretical evolution, mainly triggered by
a strong-coupling approach to include the electron-lattice
coupling, can now describe the transition between the FMI and FMM
with varying doping.
Pai {\it et al}.\ \cite{pai} have recently proposed a new effective
low-energy Hamiltonian starting from  two qualitatively different
coexisting vibronic states at  each site of the lattice these
being labeled as $\it{l}$ and $\it{b}$.
Here $\it{l}$ vibrons describe a localized Jahn-Teller (JT) polaron
and the other $\it{b}$ vibrons form a broad and dispersive band.
Within this approach, the insulating gap in the FMI phase
at low temperature is identified as the  $T$ = 0 electrical gap $\Delta$
between the occupied $\it{l}$ levels and the unoccupied $\it{b}$ band bottom.
In principle, the theory is able to make detailed predictions on the
electronic structure and transport properties through the whole
range of compositions between the FMI and FMM phase in the
perovskite manganites.

These theoretical developments and other experimental findings
inspired us to explore the low temperature transport and specific
heat properties  of mixed-valent manganites.
The present analysis is based on experimental data obtained on a
series of single-crystals from the Pb doped NdMnO$_{3}$ system,
which is a less studied member of the perovskite manganite family\cite{kusters, ng}.
From our own earlier studies on these crystals, we presented
a tentative phase diagram of
Nd$_{1-x}$Pb$_{x}$MnO$_{3}$ for wide range of temperature
and composition \cite{ng1} and some of their
physical properties \cite{ng1,ng2,sa05,ng06}.
In the present investigation, we report further properties of
these Nd$_{1-x}$Pb$_{x}$MnO$_{3}$ single crystals and a detailed
analysis of their low temperature resistivity
and specific heat data.

\section{Experiments}
Single crystals of  Nd$_{1-x}$Pb$_{x}$MnO$_{3}$ were  grown by a
high temperature solution growth method using PbO/PbF$_{2}$ as
flux, as reported earlier \cite{ng}.
The compositional analysis has been carried out
by Energy Dispersive X-ray (EDX) and then substantiated
by Inductively Coupled Plasma Atomic Emission Spectroscopy (ICPAES).
Resistivity measurements were carried out
by the standard four probe method in the range 4-300K.
The electrical contacts were generally  made on the (100) plane of
the single crystal samples.
The contact material was a Ag-15 weight $\%$ In alloy.
Magnetoresistance (MR) was measured in similar temperature ranges
at 7~T magnetic field.
The specific heat measurements were carried
out in a physical property measurement
system (PPMS model 6000, Quantum Design) from 40~K down to 2~K.

\section{Results}\label{SectResults}

\subsection{Transport}\label{SubTransport}
Resistivity data for the Nd$_{1-x}$Pb$_{x}$MnO$_{3}$ single
crystals were already presented in Ref.~\cite{ng1} focussing on
the metal-insulator (MI) transition at intermediate temperatures
around 150~K.
%
%
%
%
%
%
%
The activated behaviour, $\rho$ = $\rho_{0}$ $\exp(E/k_{B}T)$, is
put into evidence by the logarithmic plot of the resistivity
$\rho$ as a function of inverse temperature $1/T$.
This is shown for $x$ = 0.15, 0.3, 0.4 and 0.5 in Figure~1.
The slopes of the straight
sections of the resistivity in this
plot measure the activation energy $E$.
The crystals with $x$ = 0.15 and 0.3 display an activated
behaviour in both the paramagnetic and ferromagnetic state, albeit
with different activation energies.
The activation energy $E \equiv E_{A}$ in the high  temperature
paramagnetic insulating phase can be related to transport through
Jahn-Teller(JT) polarons, as discussed in  the literature
\cite{millis}.
The activation energy $E_{A}$ depends on $x$ and becomes minimum
for the most metallic sample close to $x=0.4$ in our series of
crystals \cite{ng1}.

For the ferromagnetic insulating samples, the same type of
behaviour holds with an activation energy given by a gap energy,
$E=\Delta$, in some temperature range below $T_C$ towards zero
temperature.
The gap energy $\Delta$, as derived from activated fits, decreases
with the dopant concentration $x$.
The dependence of the gap energy  $\Delta$ versus dopant
concentration $x$ is shown in Figure~2.
It can be compared with theoretical predictions.
The critical doping for the $T = 0$ ferromagnetic insulator to
ferromagnetic metal transition is to be determined by the
vanishing of  $\Delta$.
According to the theory of Pai {\it et al}.\ \cite{pai} a dependence
$\Delta(x)=E_{JT}-x^{1/2}\,D_0$ is expected, where $E_{JT}$ is
the energy scale for the electron JT-phonon coupling, and $D_0$ is
a measure of the electronic band-width of the itinerant $e_g$
electrons of Mn.
Fits for the doping dependence of the experimental values of
$\Delta(x)$ to this square-root behaviour are shown in Figure~2.
The values  obtained from this fit for $E_{JT}$ and  $D_{0}$  are
around 273 ($\pm 38$) and 493 ($\pm79 $)  meV
respectively, when fitting  was carried out in $\it{x}$= 0.15-0.3 range.
On the other hand, the fitting in $x$= 0.2-0.3 range yields $E_{JT}$
and  $D_{0}$   as   around 195 ($\pm 12$) and 339( $\pm
25$) meV respectively.
%
It is observed that  the system for dopant concentration $x$ =
0.15 is not well fit by the theoretical square-root dependence for
$\Delta(x)$.
As seen from Figure~2 the experimentally determined value for
$\Delta(x=0.15)$ is appreciably increased.
It is probable that at this low doping level further effects
increase the apparent activation energy for the electronic
transport.
The extrapolation of $\Delta(x)$ towards a vanishing gap
$\Delta=0$ yields a critical doping level
for the zero-temperature MI transition.
The critical concentration for the zero temperature MI transition
in this system has been calculated from the parameters obtained
from the fitting  excluding the point
corresponding to $x = 0.15$  giving a value $\it{x_{c}}$  = ${(E_{JT}/D_{0})^{2}}$ = 0.33($\pm 0.02$).
We consider that the fit excluding $x$ = 0.15 data provides the
more reliable estimate of the critical concentration.

Now, assuming that $\Delta$ vanishes for this critical doping
level $x=x_c$, we can proceed to analyze the transport data for
samples with $x$ = 0.4, 0.42, 0.5 $ > x_c$, which display
metallic character at low temperatures.
Essentially, $\it{x}$= 0.4 and 0.42 show  similar behaviour.
The resistivity at low temperature can have contributions from
residual resistivity $\rho_0$ due to static defects independent
on temperature and from scattering by elementary excitations.
The effect of electron-electron scattering can be described by a
term proportional to  $T^2$ \cite{jaime}.
Taking  these two contributions into account
the resistivity should follow the relationship as $\rho$ =
$\rho_{0}$ + $A T^{2}$.
The data for $x$ = 0.42 (Figure~3(a)) are expected
to follow this conventional behaviour.
However, the $\rho(T)$ dependence with $\it{x}$= 0.5 sample shows
a shallow but distinct minimum
in the low-temperature region (see inset in Figure~1)
indicating further effects.
The resistivity  minimum persists in the presence
of magnetic fields and even becomes deeper,
while shifting to higher temperature of
about 110~K in a 5 Tesla field (Figure~3(b)).
We assume that this anomaly is related to weak localization and
electron-electron interaction in the correlated electron system
owing to static disorder and large Coulomb interactions \cite{Lee}.
Additionally, higher order contributions in temperature may play a role.
These are two-magnon interactions with a term $\propto T^{4.5}$ and an
electron -phonon (e-p) term $\propto T^5$.
Because it is difficult to distinguish between these two terms,
we do not consider these two possible contributions separately,
rather we restrict our fits to the two-magnon term only.
Hence, the resistivity for the metallic low-temperature range
should be described by
\begin{eqnarray}
\rho(T) = \rho_{0}+ A T^{2} + \rho_{\epsilon} T^{1/2} + \rho_{m} T^{4.5}\,.
\end{eqnarray}

For the resistivity of the $x$ = 0.42 sample,
no indication of weak localization could be observed.
To verify this, we have plotted the resistivity data
after subtracting the residual resistivity against $T^{1/2}$
(see inset of Figure~3(a) ).
The data are not linear as a function of  $T^{1/2}$
in that temperature range, whereas linearity is
supposed to be a prominent signature in support of weak
localization and a $T^{1/2}$-term due to  e-e interactions \cite{ziese,mich}.
Hence, resistivity  data for $\it{x}$ = 0.42  have
been fitted according to equation~(1) without
the $T^{1/2}$ term. The result is shown in Figure~3(a).
The fit parameters for the stable fit in the temperature range $T<50$~K
are listed in table~1.
\begin{table}[tbh]
\caption{ Summary of fitting of the resistivity data for
the $\it{x}$ = 0.42 single crystal in the low temperature  range  according to equation.~(1)
with error bars and standard deviations ($\chi^{2}$).
}
%
\begin{tabular}{ |c|c|c|c|c|c| } \hline
$\it{x}$ values & $\rho_{0}$   & $\it{A}$  & $\rho_{m}$ & T & $\chi^{2}$ \\
            &    ohm-cm  &  ohm-cm K$^{-2}$  & ohm-cm K$^{-4.5}$ &  K &  \\ \hline



 $\it{x}$ =0.42  & 0.2001  &  7.4 $\times$10$^{-6}$ &
 6.5 $\times$10$^{-11}$ & $\leq$50 & 5$\times$ 10$^{-7}$\\
  & $\pm$ 0.2 $\times$10$^{-3}$    & $\pm$4 $\times$10$^{-7}$ & $\pm$ 2$\times$
 10$^{-11}$ & &\\\hline
\end{tabular}

\end{table}\\
The resistivity data of the $x=$~0.5 crystal
require inclusion of the $\rho_{\epsilon} T^{1/2}$ term
due to weak localization.
This is obvious from the linear part of
the  plots $\rho$ versus $T^{1/2}$ in the Inset of figure~3(b)
with and without applied magnetic field.
However, the presence of the electronic contribution $A T^{2}$ to resistivity
is uncertain.
As listed in Table~2 we have tried various fits for the data in zero magnetic field:
a fit in the low temperature range excluding the  $T^{4.5}$ term
in the temperature range  $T\leq$60~K
leads to an insignificantly small or even negative
contribution $A T^{2}$ with large errors.
Similar results for the $A T^{2}$ term are found for fits
by including the $T^{4.5}$ term in an
extended temperature range $T\le 100$~K.
Therefore, this electronic term appears to be irrelevant for
a valid description of the data.
However,  one can  get an upper estimate of the parameter
$A$ ( 6.3 x 10$^{-6}$ ohm cm K$^{-3/2}$) from the fit in the range $T \leq$60~K.
By putting $A\equiv 0$, we find a satisfactory fit in
the same temperature range in zero field.
A fit with similar quality is achieved
for the resistivity data in a field of 5~Tesla (table~2 and figure~3(b).
Here, the magnitude of  $\rho_{0}$ is slightly  increased
as compared to the fit to the zero-field resistivity.
It is to be noted that the lowest possible temperature,
at which we could measure the resistivity is around 5~K for $x$ = 0.5.
Hence, the value $\rho_{0}$ is actually close to $\rho_{5K}$.
In the fits, there is a trade-off between the negative $T^{1/2}$ term
and the $\rho_0$,
which leads to some systematic interrelation between
these contributions.
Therefore, the shift in $\rho_{0}$ is probably
not related to a real physical effect due to magnetic field.
However, the strong increase in the magnitude of $\rho_{\varepsilon}$
due to the magnetic field is clear, see figure~3(b).
The two-magnon term $T^{4.5}$ is suppressed in a magnetic field,
which suggests that it is dominated by
two-magnon scattering contributions
that are subdued by the magnetic field.
\begin{table}[tbh]
\caption{Summary for fits of the resistivity data
for the sample
with $\it{x}$ at $\it{\mu_{0}H}$ = 0 and 5T
in different temperature ranges ($\chi^{2}$ is
the  standard deviation)}.
\begin{tabular}{ |c|c|c|c|c|c|c| } \hline
$\it{x}$,$\it{H}$ values & $\rho_{0}$ & $\it{A}$ &  $\rho_{\epsilon}$   & $\rho_{m}$  & $\it{T}$ & $\chi^{2}$ \\
           &    ohm-cm & ohm-cm K$^{-2}$  & ohm-cm K$^{-1/2}$ &   ohm-cm K$^{-4.5}$ & K &\\ \hline

  $\it{x}$=0.5 & 0.626 & -6.2 $\times$10$^{-7}$ & -1 $\times$ 10$^{-2}$ & ... & $\leq$40  & 3.4
                $\times$10$^{-7}$\\
  $\it{H}$=0  & $\pm$0.1 $\times$ 10$^{-2}$ & $\pm$9.34 $\times$10$^{-7}$& $\pm$0.3 $\times$
                10$^{-3}$& & &\\
                & 0.6327& 6.3 $\times$10$^{-6}$ & -1 $\times$ 10$^{-2}$ &... &$\leq$60 & 2.4 $\times$10$^{-6}$\\
                &$\pm$0.1 $\times$ 10$^{-3}$ & $\pm$7.4$\times$10$^{-7}$ & $\pm$ 0.5 $\times$10$^{-3}$ & & &\\
                &0.6296 & 5.26 $\times$ 10$^{-7}$& -1 $\times$ 10$^{-2}$ &  1.6$\times$10$^{-10}$& $\leq$100 & 2 $\times$10$^{-6}$\\
                & $\pm$0.1$\times$ 10$^{-3}$ & 7.4 $\times$ 10$^{-7}$ & $\pm$0.4 $\times$10$^{-3}$ & $\pm$6 $\times$ 10$^{-12}$& & \\

                &  0.6287 &... &-1 $\times$ 10$^{-2}$  & 1.6 $\times$10 $^{-10}$& $\leq$100& 2 $\times$10$^{-6}$\\

                & $\pm$ 0.12$\times$ 10$^{-2}$& & $\pm$0.2 $\times$10$^{-3}$ &   $\pm$2$\times$10$^{-12}$& &\\

 $\it{x}$= 0.5  & 0.6513 &... & -3 $\times$10$^{-2}$   & 2.8 $\times$10$^{-11}$& $\leq$180 & 2 $\times$10$^{-5}$\\

$\it{H}$=5T&  $\pm$0.18 $\times$10$^{-2}$& & $\pm$0.2 $\times$10$^{-3}$ &  $\pm$ 2
$\times$10$^{-13}$ &  &\\\hline
\end{tabular}
\end{table}

To ascertain the essential absence of the expected electronic contribution
to the resistivity $AT^2$, we have additionally used graphical methods.
We plot the resistivity against $T^{2}$ (after subtraction of the residual resistivity).
The data for  $x=$~0.5  do not follow a straight line (inset~(b), Figure~4),
unlike those for $x=$~0.42( inset~(a), figure~4).
In order to better display the deviations from the $T^{2}$
contribution, we also have numerically differentiated the data
and plotted as  $A^{-1}$ $d\rho$/$dT^{2}$  vs $T$ in figure~4.
Here, we have  used the calculated  value of $A$  from the slopes of
the straight line fits (see insets, figure~4).
In the plot of figure~4, the data show an upward deviation from 1 at
higher temperatures. This should be related to the expected weak
two-magnon scattering ( $T^{4.5}$) and  electron-phonon scattering
($T^{5}$) contributions.

\subsection{Specific heat}\label{SubC}
The specific heat $C$ vs temperature data for the series
of Nd$_{1-x}$Pb$_{x}$MnO$_{3}$ crystals are plotted in Figure~5.
For  manganites the basic contributions to the specific heat in
the low temperature range can be described by \cite{bf}.
\begin{eqnarray}
\label{specheat} C = \beta_{3/2} T^{3/2} +  \gamma T  +  B_{3}
T^{3} +  B_{5} T^{5}.
\end{eqnarray}
Here, $\beta_{3/2}$ is the coefficient of the contribution from
spin wave excitations for ferromagnetic order, $\gamma$ is the
coefficient of the electronic specific heat, $B_{3}$ and  $B_{5}$
are coefficients of the contribution from the lattice.
In the  temperature range $T<$~15~K, an additional contribution
from a Schottky-like effect strongly influences the behaviour of
the specific heat for the Nd-based systems \cite{ng2}.
The best range for the fitting according to
Eq.~(\ref{specheat}) has been found to be 20 --- 40K.
For this temperature range nuclear hyperfine effects in the
specific heat need not be considered, as they contribute
appreciably only at much lower temperature $ T <$~2~K.
The results of the fitting by Eq.~(\ref{specheat}) is
shown for $x$ = 0.4 as a representative plot in Figure~6.
Similar fits have been achieved for $x$ = 0.15
and 0.5 (not shown in the figure).
An analysis of the specific heat of the $x$=0.3 crystals has
already been reported in Ref.~\cite{ng2}.

\begin{table}[tbh]
\caption{%
Summary of fitting for the specific heat data
in the temperature  range 20--40K. The values of the spin-stiffness $D$
derived from magnetization data via Bloch $T^{3/2}$ law have been used to
fix the  $\beta_{3/2}$ parameter for the ferromagnetic spin-wave contribution.
The definition of the other coefficients are given in the text. $\chi^{2}$ is
the  standard deviation. The parameters for $x$ = 0.3 are taken from Ref.~\cite{ng2}
}

%
\begin{tabular}{ |c|c|c|c|c|c|c| } \hline
Composition & $\beta_{3/2}$ &  $\gamma$  &  $B_{3}$     &  $B_{5}$  & $D$  & $\chi^{2}$\\
   & mJ mole$^{-1}$   &  mJ mole$^{-1}$  & mJ mole$^{-1}$   &  mJ mole$^{-1}$ & meV \\
   &  K$^{-5/2}$  & K$^{-2}$ & K$^{-4}$ & K$^{-6}$ & \AA$^{2}$ &\\

   & &   &    &   $\times$ 10$^{-3}$ & &\\ \hline


Nd$_{0.85}$Pb$_{0.15}$MnO$_{3}$  & 19.57 &  71.82 &  0.26 &  -0.068 &
17.3 & 4.8 $\times$10$^{-3}$ \\
  & ...&  $\pm$0.11 &  $\pm$0.01  &  $\pm$0.004 &
.. &  \\
Nd$_{0.7}$Pb$_{0.3}$MnO$_{3}$  & 6.51&  71.73 & 0.356 &
-0.98 & 34.9 & 1 $\times$10$^{-3}$\\
& ...&  $\pm$0.13 &  $\pm$0.05  &  $\pm$0.007 &
.. &  \\
 Nd$_{0.6}$Pb$_{0.4}$MnO$_{3}$  & 6 &  68.64 & 0.387
&  -0.112  &  37.8& 3.3 $\times$10$^{-3}$\\
& ...&  $\pm$0.01 &  $\pm$0.004  &  $\pm$0.002 &
.. &  \\
Nd$_{0.5}$Pb$_{0.5}$MnO$_{3}$  & 1 & 111.51  & 0.411 & -0.125
&  125.8 & 6 $\times$10$^{-3}$\\
& ...&  $\pm$0.02 &  $\pm$0.005  &  $\pm$0.003 &
.. &  \\ \hline

\end{tabular}
\end{table}

The detailed results of fitting are given in Table~3.
We have extracted the initial value of $\gamma$ from the $y$-axis intercept
in the  plot of $C/T$ vs $T^{2}$ (inset of figure~6).
To determine the magnon contribution to the specific heat, we
extracted the  spin stiffness constants $D$ from the magnetization
data \cite{ng2}
using Bloch  $T^{3/2}$ law (see table~3).
Subsequently, the corresponding values of $\beta_{3/2}$ are
calculated from $D$  using the relation $\beta_{3/2}$ = 0.113 $R
a^{3}$ ($k_{B}/D$)$^{3/2}$. Here,  $R$ = 8.314 J K$^{-1}$
mole$^{-1}$ is the universal gas constant and $a$ is the lattice
parameter of the elementary perovskite cell \cite{ng2}.
Then the remaining coefficients  $B_{3}$ and $B_{5}$ have been determined by fitting.
The coefficient $B_3$ corresponds to the  Debye contribution to
the  specific heat at low temperature which can be expressed as,
\begin{eqnarray}
C_{\mathrm{Debye}} = (12/5)\ r R
\pi^{4}\left(\frac{T}{\theta_{\mathrm{D}}}\right)^{3} \
\end{eqnarray}
where $r$ is number of atoms in the unit cell, i.e. $r=5$, $R$ the
universal gas constant, and $\theta_{\mathrm{D}}$ is the Debye
temperature\cite{esr}.
We have calculated the values of $\theta_{D}$ for $x$ = 0.15, 0.3,
0.4 and 0.5 and they are 332~K, 292~K, 291~K  and 285~K, respectively.
Generally, the  electronic specific heat term is not expected to
be present for samples  with $x$ = 0.15, 0.3 which are insulators
at low temperature.
However, it is observed that the magnitude of $\gamma$ is
unusually large for all these samples. The enhanced values of
$\gamma$  are most probably not related to the conduction electrons.
The corresponding specific energy contribution may be due to
magnetic effects related to the Nd ions and the Mn-sublattice \cite{ng2}.
In particular, the  magnitude of $\gamma$  is presumably influenced
by the tail of the  Schottky-like anomaly due to the presence of Nd ions.

As can be seen from Figure~5, the specific heat below 15~K  has
a strong Schottky-like anomaly for all  the samples.
This effect is due to the Zeeman-like splitting of crystal-field
ground state multiplets in the Nd$^{3+}$ ions, which was already
described for Nd$_{0.67}$Sr$_{0.33}$MnO$_{3}$ by Gordon
{\it et al}.\ \cite{gor}.
A similar phenomenon was reported for
Pr$_{0.8}$Sr$_{0.2}$MnO$_{3}$ \cite{wahl}.
Nd$^{3+}$ ions have a ten-fold degeneracy for the ground-state $J$
multiplet $^{4}I_{9/2}$ which is split by the crystal-field into
five Kramers Doublets \cite{podle}.
An  effective  molecular field $H_{mf}$ is assumed to be present
at Nd sites.
Although this  field splits each of the  five crystal-field
doublets, at low temperature, only the ground state doublet needs
to be considered \cite{gor}.
Assuming that the effective moment of Nd$^{3+}$ ions in the ground
state to be $\mu_{Nd}$ and that the splitting of the doublet is
$\Delta_s= 2\mu_{Nd}\,H_{mf}$, a contribution from a two-level
Schottky function should fit the excess specific heat at low
temperature ($T<$15~K).
We have  used the Schottky function for a two-level system as
\cite{esr,wahl}
\begin{eqnarray}
C_{\mathrm{Sch}}(T,H)= n_{\mathrm{Sch}} N_a k_{\mathrm{B}}
\left(\frac{\Delta_s}{k_{\mathrm{B}}T}\right)^{2}
\left[\frac{\exp(\frac{\Delta_s}{k_{\mathrm{B}}T})} {(1+
\exp(\frac{\Delta_s}{k_{\mathrm{B}}T}))^{2}}\right]
\end{eqnarray}
where $n_{\mathrm{Sch}}$ is the coefficient of the contribution
from the Schottky effect, and $N_a$ is the Avogadro number.
The fitting is carried out by adding this Schottky term to
Equation (3), while keeping the lattice contribution (parameters
$B_3$, $B_5$) and the spin-wave contribution ($\beta_{3/2}$)
fixed.
The result is shown in figure~7.
The values of  Schottky gaps $\Delta_{s}$ and Schottky co-efficients
$n_{\mathrm{Sch}}$ for $x$ = 0.15, 0.3, 0.4, 0.5
obtained by fitting  are  1.14, 0.95, 0.88, 0.68 meV and 0.5, 0.54, 0.56, 0.34 respectively.
It has been noticed that this anomaly can be better fitted
by a modification of the linear contribution $\gamma T$ in the
temperature range 2 to 15~K .
%
%
For example, if we relax the value of $\gamma$ during fitting, the
modified $\gamma$ values for $x$ = 0.15, 0.3, 0.4 and 0.5 will
become   44, 37, 29 and 61 mJ mole$^{-1}$ K$^{-2}$ respectively.
Thus, the  fact that  the enhanced linear contribution $\gamma T$
to the specific heat (as shown in table~ 2) is affected by
magnetic contributions is reflected by these reduced $\gamma$
values \cite{ng2}.
We assume this broad linear specific heat contribution  to
originate from possible ordering of Nd moments.
We observe that the magnitude
of the Schottky gap $\Delta_{s}$
increases with the concentration of Nd$^{3+}$ (figure~8) ions.
Equivalently, the  molecular field experienced by Nd ion is  the
strongest at the lowest $x$  and decreases as $x$ increases.
However, the expected full contribution
of the split ground-state Kramers doublet from the Nd$^{3+}$-ions is not found
in this fit, which would require that $n_{Sch}=1-x$.
Also the fit in the low temperature range is not overall satisfactory
as seen in figure~7.
\section{Discussion}\label{SectDisc}
We  have seen that the gap energy $\Delta$ in  the
ferromagnetic insulating phase at low temperature varies with $x$
and vanishes at a critical value  $x_{c}$ $\approx$ 0.33.
Although the origin of the gap is not clearly understood yet, we
can find some explanation
in the light of the theory  by Pai {\it et al}.\ \cite{pai}.
The theory considers three important on-site interactions in
manganites, namely the Jahn-Teller (JT) effect, Hund's rule
coupling ($J_{H}$), and  Coulomb repulsion $U$.
The theory is based on the new idea of co-existing localized JT
polaronic ($\it{l}$) and broad band ($\it{b}$) itinerant electrons
from e$_{g}$ states.
The  effective bandwidth $2W$ ($W=x^{1/2} D_{0}$) of the $\it{b}$
band decreases significantly as $x$ decreases for any sizable $U$.
Consequently, the bottom of the $\it{b}$ band shifts above the
Fermi level for small $x$.
At $T$ = 0 all e$_{g}$ electrons become localized as
$\it{l}$ polarons.
Mobile $\it{b}$ states are occupied only by thermal excitations
across the gap.
The system is still ferromagnetic because of the Hund's rule
coupling $J_{H}$ that remains operative also for  $\it{l}$
polarons.
This explains the insulating ferromagnetic behaviour at low doping
with a thermally activated electronic transport \cite{rama04}.
Furthermore, with the increasing  $x$, $W$ increases and beyond a
critical concentration  $x_{c}$, when the band width equals the
JT-distortion energy, $W(x) = E_{JT}$, the low
temperature state becomes a ferromagnetic metal.

We have found the  values for  $D_{0}$ and $E_{JT}$ of
Nd$_{1-x}$Pb$_{x}$MnO$_{3}$ from our analysis.
These are of acceptable order of magnitude,
but appreciably smaller than
the  values estimated  by Pai {\it et al}.\ .
The smaller value for $D_{0}$ may be related to
the decrease of  half-bandwidth
due to smaller cationic radius
in Nd$_{1-x}$Pb$_{x}$MnO$_{3}$ which is
comparatively smaller than in the wide-band
systems like  La$_{1-x}$Sr$_{x}$MnO$_{3}$ \cite{pai}.
However, the measured activation energy $\Delta(x)$ for  the
composition Nd$_{0.85}$Pb$_{0.15}$MnO$_{3}$  does not fit well
with the dependence expected by the theory of Pai $et al$.\ \cite{pai}.
This is possibly related to appreciable antiferromagnetic
couplings giving rise to stronger spin-disorder scattering
at this low doping close to the insulating antiferromagnetic phase.
This could cause an increased apparent activation energy
for the crystal with $x=0.15$.
It is noticed that the metallic phase sets in above $x$ = 0.3 and
for compositions beyond this level
we have found a clear metal-insulator transition at
high temperatures.

The empirical relationship between the coefficient
$A$ for the electronic contribution to resistivity
and the coefficient $\gamma$ of
the electronic specific heat
has been found by  Kadowaki and Woods
with $A/\gamma^{2}$ $\approx$ 1 $\times$ 10$^{-5}$  $\mu$
$\Omega$ cm (mole K $^{2}$/mJ)$^{2}$  \cite{kadawaki}.
For the metallic composition $x=0.42$  we can use
the determined values for $A$ and $\gamma$ to estimate
this ratio.
We find a Kadowaki-Woods ratio around  850 x 10$^{-5}$
$\mu$ $\Omega$ cm (mole K$^{2}$/mJ)$^{2}$.
This means that the properties of the metal-like manganites are far
from those expected for a normal metal.
Since  the Kadowaki-Woods empirical relationship is only valid for
pure metals, it is not strictly applicable to the present system.
However, considering  the fact that  samples with $x$ $\geq$ 0.4 are
metallic at low temperature we can   assume that there should be a
positive contribution  from e-e scattering.
In fact, we have found that  there is  still a positive
contribution from e-e interaction  with  conventional exponent $\propto$ $T^{2}$ in $x$ = 0.42.
But, this contribution is not prominent
in $x$ = 0.5 at $\mu_{0}H$ =0, possibly due to
the resistivity minimum found for this composition.
We have attributed this resistivity minimum to weak localization
in a disordered correlated system \cite{Lee}.
Resistivity minima at low temperature are also a typical
characteristic of the Kondo effect \cite{Kondo}.
However, the resistivity minima due to a Kondo effect should
disappear in relatively weak applied fields,
which is not observed in the present case.
Moreover, the manganite system at $x=0.5$ behaves as a homogeneous
ferromagnetic metal and not like a metallic alloy with dilute
magnetic impurities.
Hence, the resistivity minimum observed here for the $x=0.5$ single crystal
should not be related to a Kondo effect.

On the other hand, the enhanced weak localization contribution
$\rho_{\epsilon}$ term in the presence of  magnetic field is
an expected effect.
According to Ref.~\cite{Lee}, the e-e interaction in the
presence of strong disorder is special and the magnetic field
enhances this interaction which results in an increased resistivity.
From the resistivity in the $x$=0.5 sample, we also find
a reduction in the $\rho_{p}$ term by a magnetic field,
which is consistent with a reduced two-magnon scattering.
Here, one should  remember  that it has
not been possible to distinguish the $T^{4.5}$ term due to two-magnon scattering
from the $T^{5}$ electron-phonon interaction term.

The weak localization and $T^{1/2}$ e-e interaction should
be attributed to the static disorder in these manganites
caused by the mixed A-site substitution of tri-valent rare-earth Nd$^{3+}$
by divalent ions (Pb$^{2+}$), which gives rise to some static electronic effects.
The magnitude of disorder  can be quantified by the A-site variance
$\sigma^{2}=<r^{2}_{A}>-<r_{A}>^{2}$  where
$r_{A}$ is average cationic radius \cite{cnr}.
The $\sigma^{2}$ for $x$ = 0.3 is 0.0101 \cite{ng2}.
Taking this value into consideration
the disorder for $x$ = 0.42 and 0.5 is around 0.0141 and 0.0168 respectively.
The difference between the fitting of  resistivity
data for $x$ = 0.42 and 0.5 (see table~1 and 2) is clearly observed.
Hence, the disorder may not be large enough to cause
a prominent effect of weak localization in $x$ = 0.42, whereas
its influence is strong in  the sample with $x$ = 0.5.
In addition, it should be mentioned
for the specific Nd$_{1-x}$Pb$_{x}$MnO$_{3}$ system, that
the half-doped system does not show indications
of a charge-ordered/orbital-ordered and antiferromagnetic behavior.
This is consistent with the proposed phase diagram of Nd$_{1-x}$Pb$_{x}$MnO$_{3}$,
where no indication of a  charge-ordered/orbital-ordered phase
was observed around  50:50 composition\cite{ng1}.

The analysis of the specific heat data points to some particular
effects in the magnetic system across the series of crystals.
It is seen in table~3 that the value of $D$  increases with $x$,
because the average cationic radius ($<r_{A}>$)  increases with $x$ too.
Since $D$ is directly proportional to the exchange
integral, the higher $D$ value implies larger exchange coupling
resulting in higher $T_{C}$ \cite{ng2}.
The fact that the $x$ = 0.5 sample  has  the highest  $T_{C}$ supports
the above explanations.
The Debye temperature $\theta_{D}$ decreases
with the increase of $x$ here (see table~3).
However,the Debye temperatures
are smaller than usual values
reported for manganites \cite{akr}.
This may be due to the fact that the analysis has not been carried
out in the constant $\theta_{D}$ region and the value is
affected by systematic drifts \cite{ ng2, esr}.
The electronic contribution $\gamma T$ to the specific
heat is unusually large
and present even in the insulating phase
at low temperature.
We believe that this broad specific heat contribution is related
to a magnetic ordering of Nd moments, and/or to frustrated glassy
ordering in the coupled Nd-Mn magnetic system owing to the
dilution of the rare-earth A-site.
Such an effect would be poorly emulated by an enhanced value of
$\gamma$ over the temperature range up to about 30~K.
Hence, an exact evaluation of Kadowaki-Woods ratio is difficult
for two specific reasons:
(i) Even where the $A T^2$ term is
clearly present as in $x$ =0.42, the $\gamma$
from specific heat is overestimated by the magnetic
contributions to the low temperature specific heat.
Therefore, there is no clear way to estimate the Kadowaki-Woods ratio
even in that case.
In this respect, the true
electronic $\gamma$  may be smaller, thus the calculated
K-W ratio is a lower estimate.
(ii) As our study has suggested for $x$=0.5
the $A T^2$ term is not discernible in some cases.
Thus, the material does not behave like a metallic system
at all, where such a term in resistivity
would result from e-e scattering.

The specific heat measurements
at low temperature show a strong
influence of a Schottky-like anomaly.
The peaks due to Schottky-like anomaly  are significantly broadened.
This may the effect of a distribution of ground-state
splittings on the Nd due to an inhomogeneous molecular field,
e.g.  related to a Nd-Nd magnetic exchange
and the dilution on the A-site.
In particular, there may be a greater number of Nd-sites with
very low molecular field and correspondingly low Schottky-like
contribution to specific heat. This can explain why only
a part of the expected contribution of the two-level systems
to the specific heat is found in our fitting.
However, efforts to fit the Schottky-anomaly
with more than one doublet energy to mimick a distribution
of molecular fields were inconclusive owing to the great
number of necessary parameters in such fittings.
The molecular field leading to the Zeeman-split levels of
the ground state doublet in Nd$^{3+}$
includes also contributions from the Nd-Mn exchange couplings \cite{pattu}.
A direct coupling with the Mn-O subsystem may require some
particular form of superexchange like 4f-3d coupling.
If a Nd-Mn exchange mechanisms is to be
the most important contribution, then an inhomogeneity of
this molecular field would require an inhomogeneous ordering
of the Mn-sublattice.
On the other hand, the molecular field as measured
by the magnitude of the effective single Schottky
gap changes with Nd ion concentration.
It is not obvious, why and how the interaction between
Nd and Mn-O subsystem could yield a decreasing molecular field
with increasing doping $x$.
In particular, the internal
Mn-O exchange becomes stronger with increasing $x$ as seen
from the increasing Curie-temperature up to $x=0.5$.
This indicates that the change of the effective molecular
field $H_{mf}$ is influenced by the Nd content,
which may be due to an exchange interaction
between the Nd ions, e.g., by
a long-range indirect exchange of the RKKY-type.
In the case of Nd-Sr system it
has been  noticed that Nd-Nd interaction is
small compared to Nd-Mn interaction \cite{gor}.
Therefore, the detailed understanding of the magnetic
effects involving the diluted Nd$^{3+}$ sublattice on
the A-sites requires further theoretical analysis
of possible magnetic exchange mechanisms in manganites.

\section {Conclusions}\label{SectConcl}
The low temperature transport property and specific heat of
Nd$_{1-x}$Pb$_{x}$MnO$_{3}$  have been analyzed for $x$ = 0.15, 0.3 , 0.4, and 0.5.
In the considered temperature range, activated behaviour is found
in resistivity data for samples, which are ferromagnetic insulators.
This  activation gap $\Delta$ disappears beyond $x$ = 0.3, where the
true zero-temperature MI transition takes place
at an estimated critical concentration x$_{C}\approx$ 0.33.
A positive e-e scattering term  $AT^{2}$ describes
the leading resistivity effect in the low temperature
transport data for the metallic sample $x$ = 0.42.
However, the validity of $T^{2}$ contribution
to resistivity from e-e scattering
used in earlier fitting formula  is questioned here,
as it is not discernible for the composition $x$ = 0.5, which still
shows overall metallic properties.
A  negative $T^{1/2}$ contribution from e-e scattering  valid for
disordered systems explains a shallow minimum in the resistivity
for this $x$ = 0.5 crystal.
Low temperature specific heat data have been analyzed for
these samples.
Anomalously large values of the electronic specific
heat coefficient $\gamma$  probably originate from magnetic effects.
They could be related to a glassy magnetic ordering
on the diluted Nd-sublattice.
Indications of inhomogeneous magnetic properties are also
found from a wide and subdued Schottky-like contribution
to the specific heat from the Zeeman-split Kramers
ground-state doublets in Nd$^{3+}$-ions.
%

\subsection*{Acknowledgements}

NG thanks SFB 463 project funded by DFG for financial support during his work at IFW Dresden.
HLB thanks the CSIR, Government of India for financial support through an extramural research grant.

\subsection*{}

$^{*}$corresponding author, E-mail:  ghosh.nilotpal@gmail.com, present address: Institut f{\"u}r Experimentelle Physik II,Universit{\"a}t Leipzig, Linne Str. 3-5, 04103 Leipzig, Germany.

\clearpage
\section*{ Figures and Figure Captions}

\begin{figure}[h]
\includegraphics[width=10cm]{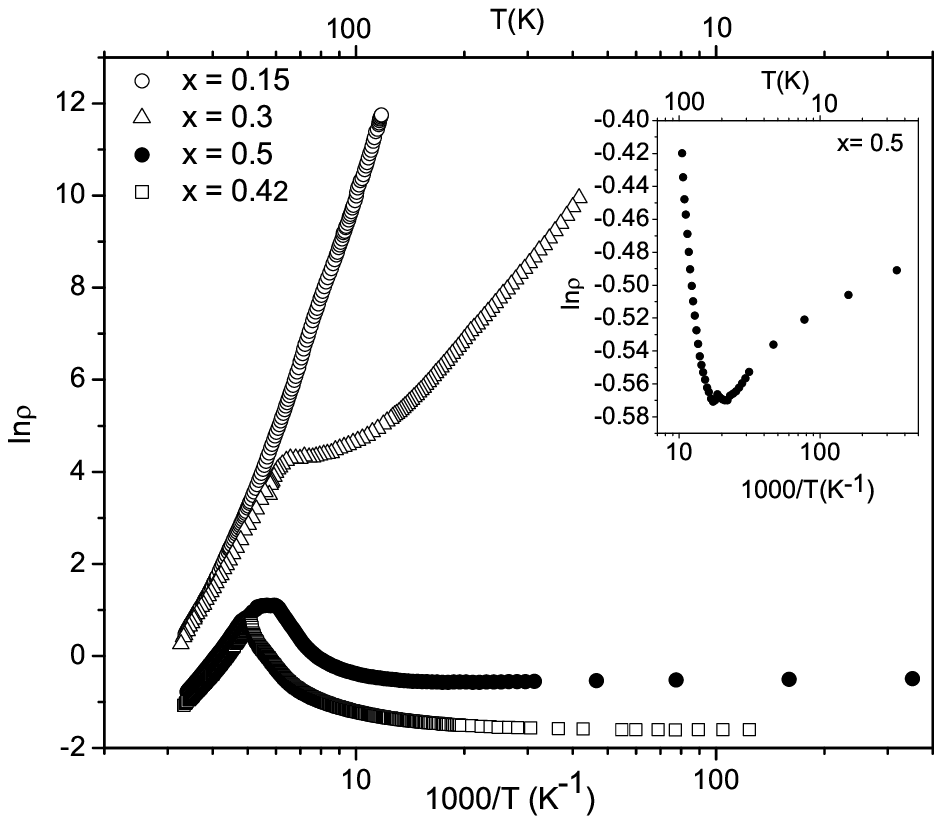}

\caption{\label{fig:epsart} The logarithmic plots of resistivity  as a function of (1000/$T$)   for  $x$ = 0.15, 0.3,  0.42 and 0.5.
Inset:  the logarithmic plot of resistivity
as a function of (1000/$T$) for $x$ = 0.5 in the low temperature range.}
\end{figure}

\clearpage

\begin{figure}
\includegraphics[width=10cm]{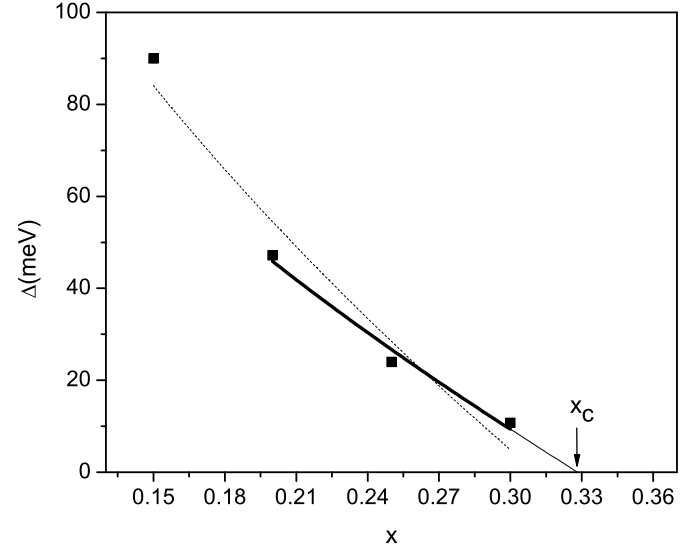}

\caption{\label{fig:epsart} The variation of  the low temperature gap $\Delta$ in the FMI phase as
a function of Pb concentration for Nd$_{1-x}$Pb$_{x}$MnO$_{3}$
for  $x$ = 0.15, 0.2, 0.25, and 0.3
(data of $x$ = 0.2 and 0.25 are taken from Ref \cite{ng1}).
The filled  square points are experimental data.
The  dashed line is the
result of a fit for the data $x$ = 0.15-0.3. The thick solid line is the result of fit for $x$ = 0.2-0.3,
which is extrapolated to the critical concentration $x_{c}$ (thin line).}

\end{figure}

 \clearpage
\begin{figure}
\includegraphics[width=10cm]{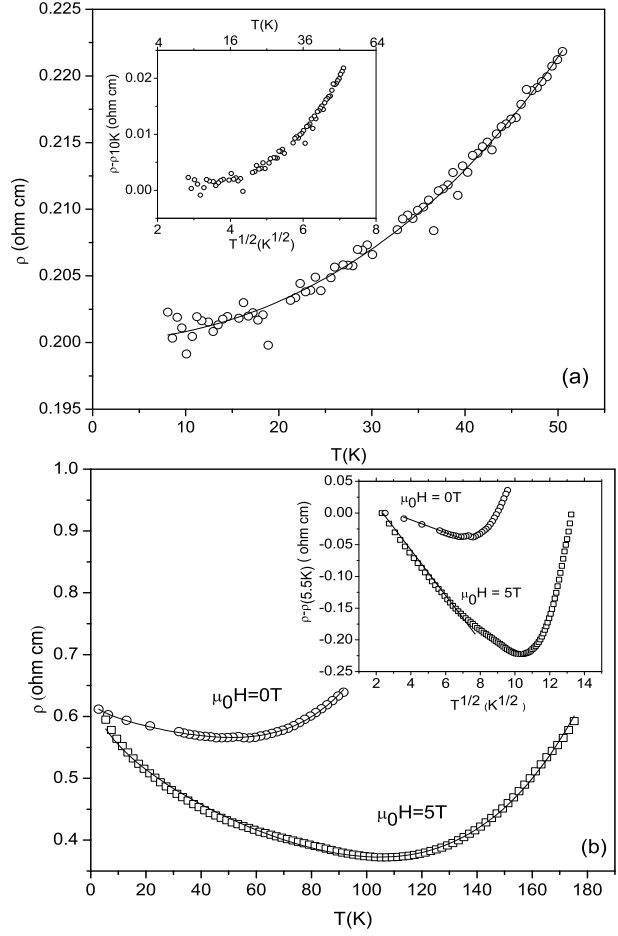}

\caption{\label{fig:epsart}(a) The  resistivity data  for $x$ = 0.42 and results of fit according to Eq.~(1)
without the $T^{1/2}$ term.
The inset shows the  plot of resistivity data
(after subtracting  the residual resistivity)
$\rho$ vs $T^{1/2}$  in zero field.
The points are experimental data and the solid lines are the fit.
(b) The  resistivity data  for $x$ = 0.5 at low temperature and  the results of the fit according to Eq.~(1)
excluding the $A\,T^{2}$ term at 0 and 5T.
The points are experimental data and solid lines are fitting results.
Inset shows the  plots of resistivity data $\rho$ vs $T^{1/2}$ for $x$ = 0.5 at 0T and 5T.
The data  in both plots below 10~K show the presence of a negative $T^{1/2}$ contribution due to e-e interaction
for $x$= 0.5 sample. The straight lines are results of linear fit.}

\end{figure}
%
%
%


%
\clearpage
\begin{figure}
 \includegraphics[width=10cm]{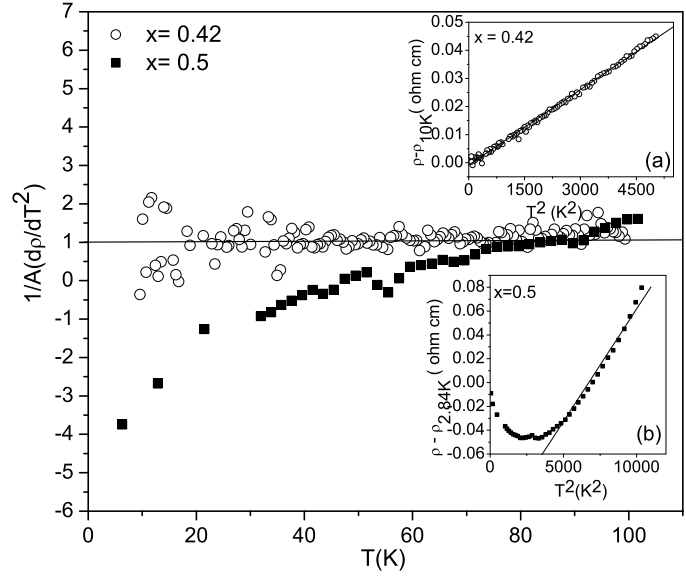}

\caption{\label{fig:epsart}The plot of  $A^{-1}$ d$\rho$/d$T^{2}$  vs $T$  for $x$ = 0.42, 0.5.
This shows the deviation from the $T^{2}$ behaviour of the
resistivity data at very low temperature for $x$ = 0.5.
The negative values of $A^{-1}$ d$\rho$/d$T^{2}$  for $x$ =
0.5  appear because there is a minimum and change of slope in
the resistivity data.
However, the data for the $x$= 0.42 sample follow the conventional $T^{2}$ behaviour.
Inset(a) and (b) show $\rho$ vs $T^{2}$ (after subtracting the residual resistivity)
for $\it{x}$ = 0.42 and 0.5 respectively. The straight lines are linear fits.}
\end{figure}

\clearpage
\begin{figure}
\includegraphics[width=10cm]{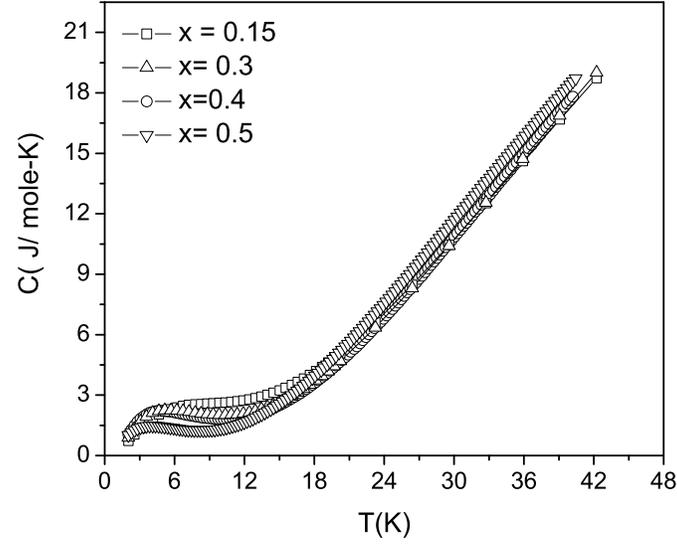}

\caption{\label{fig:epsart}The specific heat vs temperature plots of Nd$_{1-x}$Pb$_{x}$MnO$_{3}$ for  $x$ = 0.15, 0.3, 0.4 and 0.5.}

\end{figure}

\begin{figure}
 \includegraphics[width=10cm]{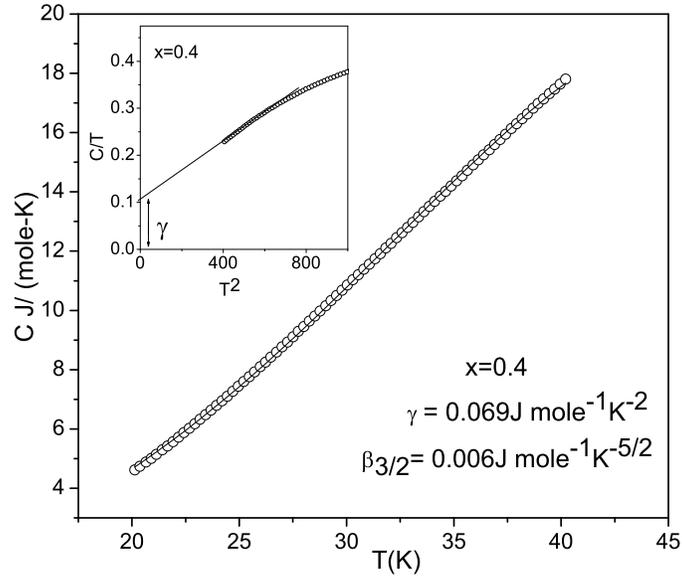}

\caption{\label{fig:epsart}Specific heat  vs temperature data for the  $x$ =  0.4 sample.
The solid line is the fit using Eq.~(2) and open circles are experimental data.
Inset shows $C/T$ vs $T^{2}$ plot for $x$=0.4} 
\end{figure}

\begin{figure}
 \includegraphics[width=10cm]{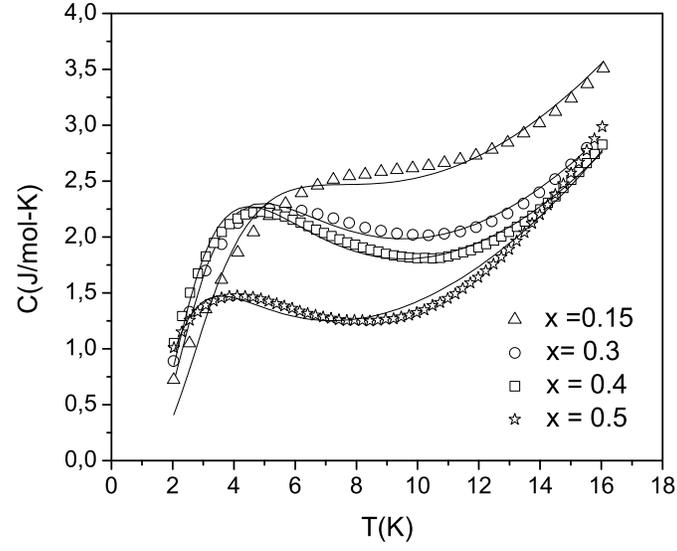}

\caption{\label{fig:epsart}Result of fits including the Schottky effect
to the specific heat data  of Nd$_{1-x}$Pb$_{x}$MnO$_{3}$
for  $x$ = 0.15, 0.3 and 0.4  in the low temperature range.
The points are experimental and the solid lines are the  fits. The data for $x$ = 0.3 have been taken from Ref.\cite{ng2}}
\end{figure}

\begin{figure}
\includegraphics[width=10cm]{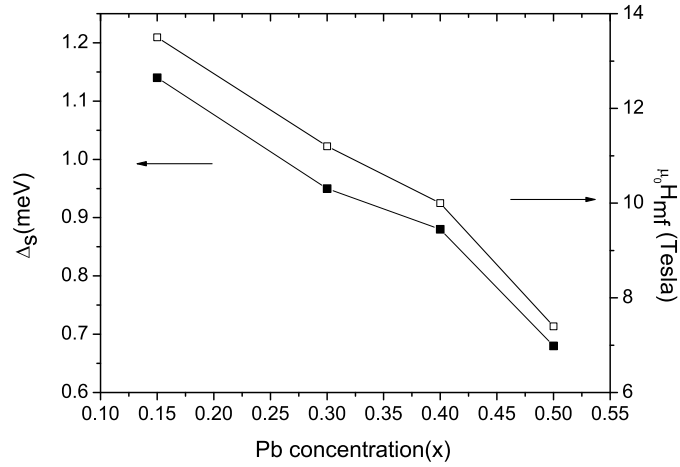}

\caption{\label{fig:epsart}The variation of the Schottky gap $\Delta_{s}$ and the molecular field
$H_{mf}$ with Pb concentration $x$ for Nd$_{1-x}$Pb$_{x}$MnO$_{3}$.}
\end{figure}

\end{document}